# Mass Extinctions by Gravitational Tides

**D. Fargion**[a,b,*]

[a] *Rome University "La Sapienza" and MIFP, Rome, Italy.*
[b] *Osservatorio Astronomico di Capodimonte, INAF, Naples, Italy*

*E-mail:* daniele.fargion@fondazione.uniroma1.it

Past and recent observations suggest that many planetary-mass or dwarf-planet objects may exist in the outer Solar System, between the Kuiper belt and the Oort cloud. Gravitational perturbations may occasionally bring some of them into the inner Solar System. The early, rare collision between the proto-Earth and the Mars-sized body Theia is generally invoked to explain the formation of the Moon. The same event probably tilted Earth's spin axis with respect to the planetary orbital plane, giving rise to the present seasons. Much more probable than a direct impact, however, are grazing or near-Earth flybys of similar or smaller objects. Such passages may have left strong tidal signatures: giant waves, large volcanic episodes, sea regressions, coherent meteor showers, and major climatic perturbations. These mechanisms could have contributed to several major biological mass extinctions over the past 600 My, as suggested by peculiar correlations in the geological record. Similar events may have occurred several times during the earlier history of Earth. Accretion of mini-planets by the Sun may also have occurred more, possibly hundred or thousand times over the last four billion years. Possibly producing additional temperature variations and climatic consequences on Earth.

*Speaker

## 1. Introduction: mass extinction , smoking gun and scapegoat

Life is short: not only individual human life but also the lifetime of many species and entire ecosystems. In the past of Earth, mass extinctions repeatedly interrupted biological evolution. In this article, we discuss a possible astrophysical mechanism for some of these events: the passage of dwarf planets or mini-planets close enough to Earth to generate extreme tidal effects. Tidal disruption is common in astrophysics. A compact object such as a white dwarf, neutron star, or black hole may tidally disrupt a companion star or a passing body, producing an accretion disk, explosive flares, or a supernova-like event. In collapsing neutron-star binaries, narrow, spinning, precessing jets may be produced. When such jets point toward Earth, satellites observe them as gamma-ray bursts (GRBs); at lower power and smaller distance, related phenomena may appear as soft gamma repeaters (SGRs). These events are very powerful. Such a astrophysical tidal catastrophe is rare near the Solar System. Here we consider a different class of tidal catastrophes: rare close passages of hidden dwarf planets from the distant Solar System.

Observed mass extinctions over the last half-billion years eliminated large fractions of animal species. In several cases, both marine and terrestrial life were rapidly affected, while amphibious animals and some birds survived more easily. These events are often associated with abrupt environmental transitions and may also correlate with large-scale geological anomalies. The extinction of dinosaurs is the best-known example. The impact of the Chicxulub asteroid , the global iridium layer, and the approximately 200 km crater in Mexico provide a successful "smoking gun" for the Cretaceous–Paleogene (K–Pg) event. However, other major extinctions did not leave such a clear crater or unique impact signature. We therefore ask whether the energy deposited by a close tidal flyby of a mini-planet could be comparable to, or even larger than, that of a large asteroid impact, while leaving no comparable crater. Such a mechanism could help explain mass extinctions without an obvious projectile or impact scar.

The frame for the present tidal Extinction Model is based on the first Newtonian understanding of the tidal force laws and-or on their deeper role by the Riemann tensor components. The expected statistical rates are, in a first approximation, based on the sane past passage of a mini-planet, Mars size, as well known, Theia in face impact on Earth. Its collision to our proto-Earth, billions of years ago, led to part of terrestrial soil and a famous relic, our present Moon.

Theia is still a hypothesized ancient planet in the early Solar System which, according to the giant-impact hypothesis, collided with the proto-Earth around 4.50 billion years ago, with some of the resulting ejected debris re-coalescing to form our Moon. Collision simulations support the idea that the two large low-shear-velocity provinces in the lower mantle of the Earth may be remnants of Theia. It is hypothesized to have been a mini-planet about the size of Mars and likely formed at the L4 or L5 Lagrange points of the Earth's orbit. Other hypotheses suggested that it may have formed in the Outer Solar System and later migrated into the Earth's orbit, and might have provided much of Earth's water. This possibility is more tuned with our model.

The Theia hit inclined the Earth spinning axis, leading to present, useful to life, seasons. Mars mini-planet or Moon size visiting objects is the base of our model. in a nearby passage. The presence of several puzzling moons or mini-moons in our Solar System is a support probe. Some of them had to be captured by large planets, as their surprising counter-rotating orbit requires. Many more show anomalous trajectories. Therefore, very far mini-planets o mini-moons might be hidden

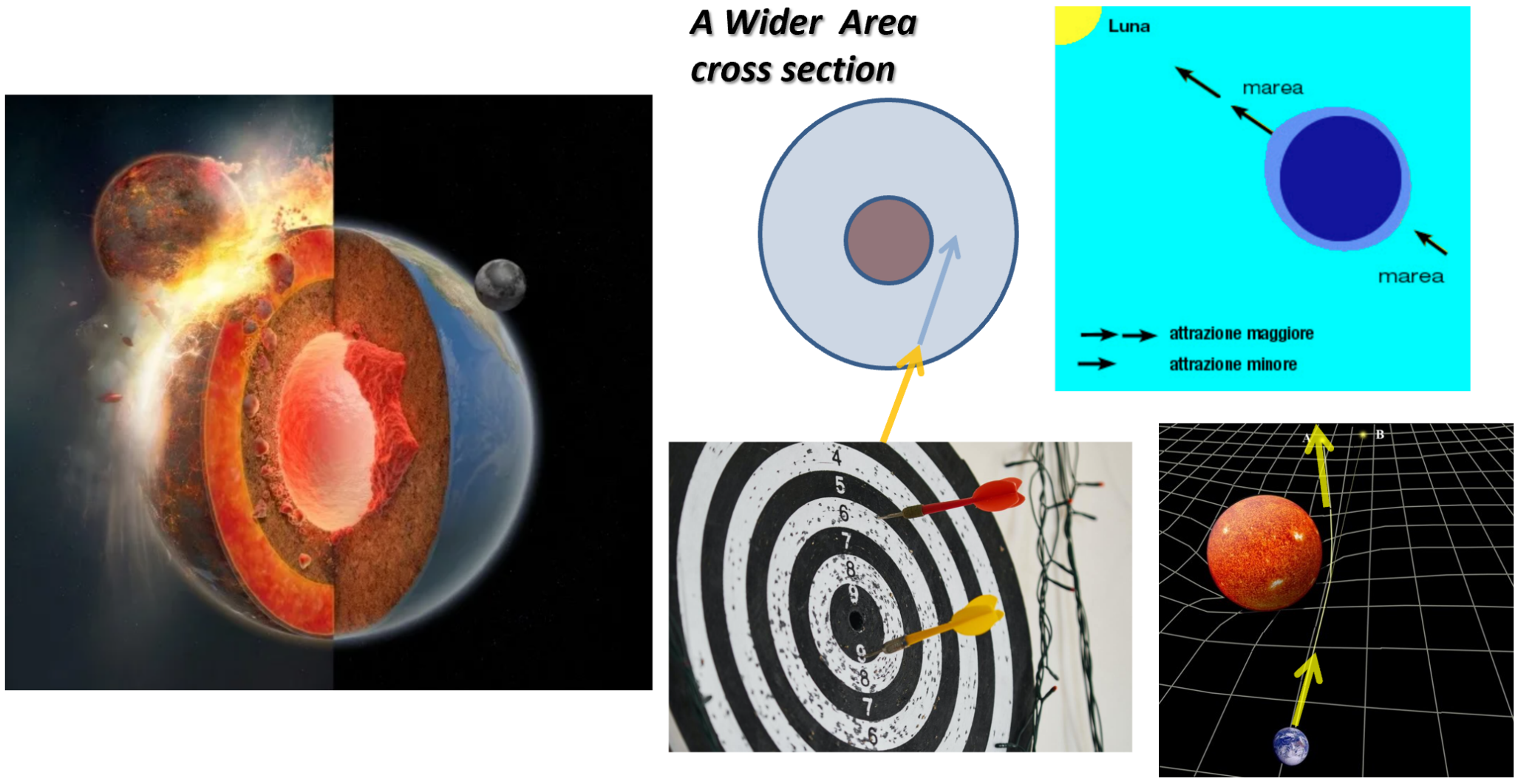

**Figure 1:** Schematic images of the hypothesized Theia impact on the proto-Earth, which led to the formation of the Moon, together with sketches of more probable grazing passages of dwarf planets.

(by their far and low luminosity) in the Kuiper belt or in a more extreme Oort cloud periphery, holding hyperbolic long-life trajectories. Their incoming in the nearer Terrestrial orbit, their flyby, or nearby deflection may occur in the past billion years. The first proposals and the recent discovery, [1], of such a far mini or dwarf planet add support and evidence to the earliest tidal extinction model [2],

The bent spin axis of many solar system objects also suggests several impacts of planets in the past. The same solar spinning axis, inclined with respect to the solar system plane (defined mainly by the Jupiter orbit), requires some past rain and collapse of hundreds or thousands of such planets or mini-planes. Obviously, a full hit is much rarer than a skimming passage on a flyby. Therefore, as an order of magnitude, the past hit of an object like Theia on Earth disk suggest a much probable fly by at wider disk area, For example, a passage at ten times the Earth radius distance. could be a hundred times more probable than a full hit. Such a Theia visit would be catastrophic even more than an asteroid impact, leading to a tidal mass extinction with no present trace or scapegoat beyond.

## Brief derivation of the Lunar Tidal Deformation

### 1. Gravitational potential of the Moon

The gravitational potential generated by the Moon at a point near the Earth's surface is

$$V = -\frac{GM_L}{|\mathbf{d} - \mathbf{r}|}$$

where $M_L$ is the lunar mass, $d$ the Earth-Moon distance, and $r$ the position on the Earth's surface.

### 2. Tidal approximation

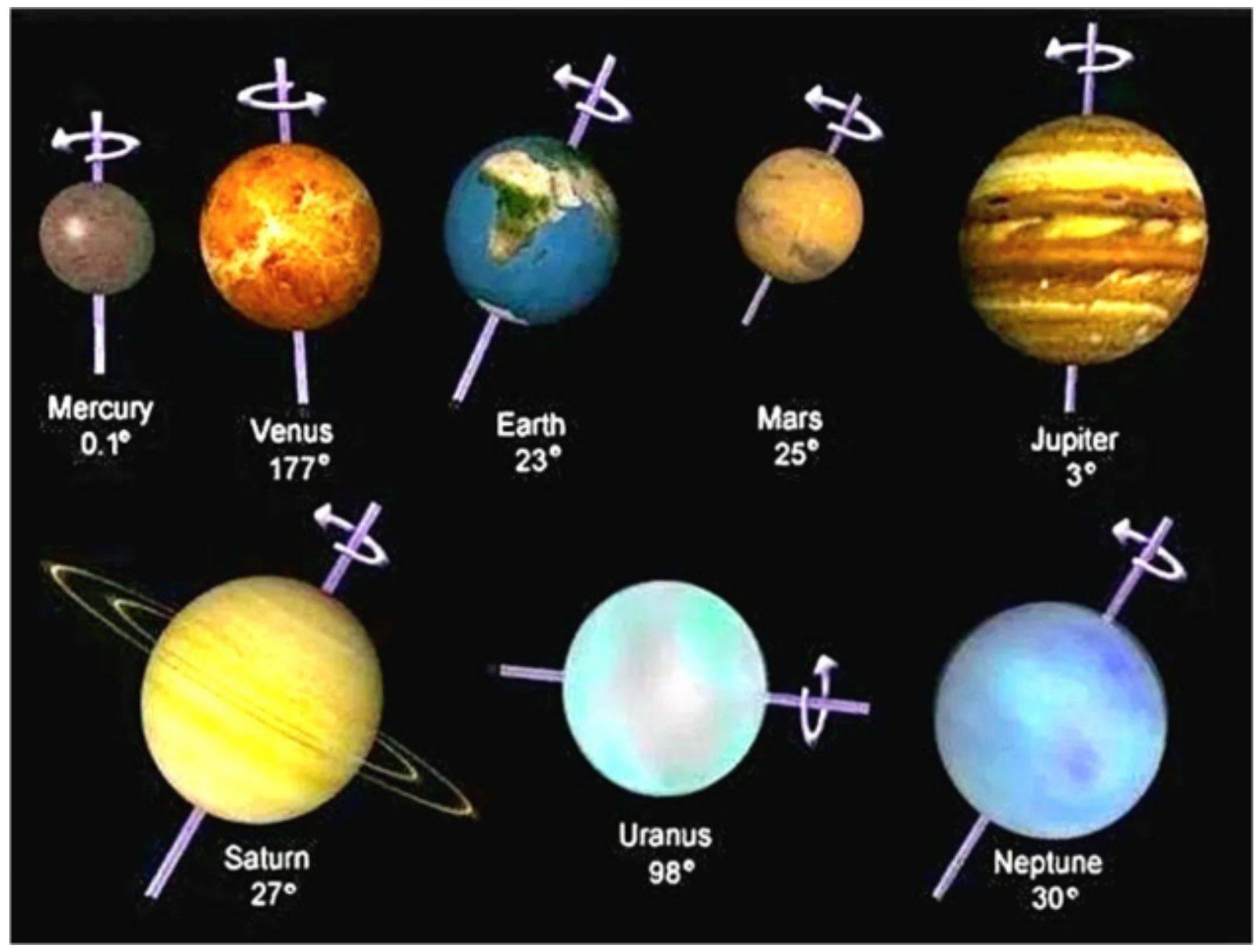

**Figure 2:** The inclined spin axis of planetary objects respect to the main solar system plane. These unexpected spinning bending should be originated by one or a few planetary collisions as for Earth and for the Uranus extreme case. axis Several random planet collisions could be necessary to change Jupiter greater angular momentum by its huge mass .

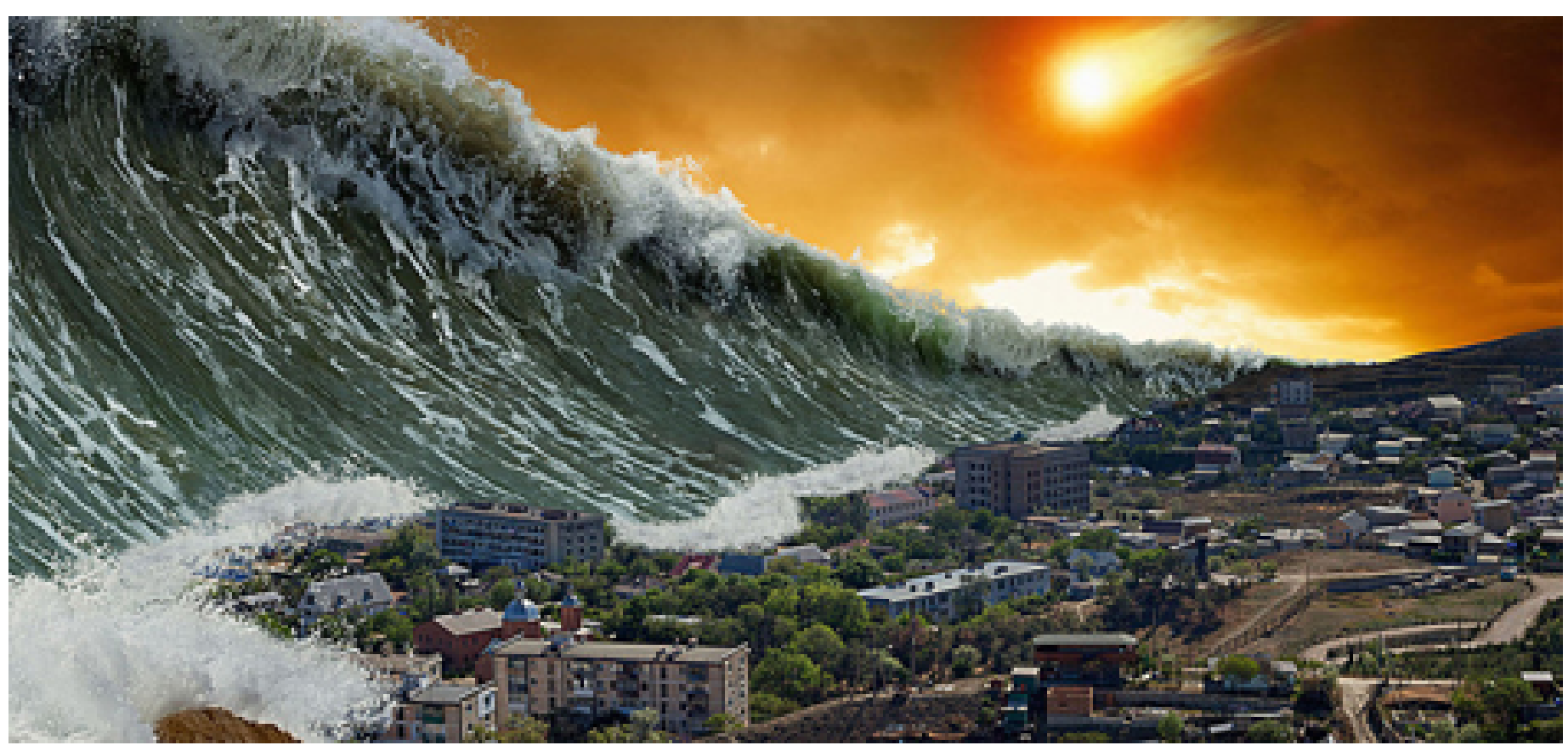

**Figure 3:** An imagine of a coast arrival of a tide tsunami in a limited event (altitude of a few tens meters). Even higher and more dramatic tidal disruptions may occur in past nearby flyby by mini-planet visit as discussed in present article.

Since $r \ll d$, we expand the potential in a series and keep terms up to second order:

$$V \approx -\frac{GM_L}{d} - \frac{GM_L}{d^2} r \cos\theta - \frac{GM_L}{2d^3} r^2 (3\cos^2\theta - 1)$$

The first term is constant, and the second produces the same acceleration on the whole Earth. The **third term** generates tides.

**3. Tidal potential**

Thus, the tidal potential is

$$U = \frac{GM_L}{2d^3} r^2 (3\cos^2\theta - 1)$$

At the surface of the Earth $r = R_T$:

$$U = \frac{GM_L R_T^2}{2d^3} (3\cos^2\theta - 1)$$

**4. Elastic response of the Earth**

The vertical deformation of the surface is proportional to the tidal potential:

$$h(\theta) = \frac{h_2}{g} U$$

Here U is the Lunar potential and g is the Earth gravity on the surface.

Substituting

$$h(\theta) = h_2 \frac{GM_L R_T^2}{gd^3} \frac{1}{2} (3\cos^2\theta - 1)$$

after an integral from $\theta = (0)$ to $\theta = (\pi/2)$ and assuming to observe tides at Earth surface:

$$\Delta h = R_T \frac{GM_L}{g} \left( \frac{R_T^2}{d^3} \right) \frac{3}{2}$$

$$\Delta h = R_T \frac{M_L}{M_T} \left( \frac{R_T}{d} \right)^3 \frac{3}{2} \quad (1)$$

Therefore, assuming the usual average Earth-Moon distance $d/R_T = 60.3$ as well as the Moon Earth ratio $\frac{M_L}{M_T} = 1/81.3$ the lunar tidal as we observe, is :

$$\Delta h = R_T \frac{1}{81.3} \left( \frac{1}{60.3} \right)^3 \frac{3}{2} \; = 0,54 \; m \quad (2)$$

This result assumes only a Moon role. The solar tide height is nearly half that of the lunar one. This lower solar tide is mostly due to the lower solar average density relative to the lunar one. Their tidal gravity terms are comparable because of the peculiar tuned distances and masses of Sun and Moon today. Their geometrical positions and their sphere radius permit a full moon-solar eclipse: Moon disk overlap almost exactly on the Sun one. The spherical masses grow as the cube of their radius . For a full eclipse, the Moon-Sun opening angle is comparable, leading to Earth-Moon and Earth-Sun distances at a ratio comparable to the Moon and Sun radius (as observed). The tides on Earth decrease with the cube of the inverse power of the distance from the object, but the tides are also proportional to the mass of the object (related to the cube of their radius) : in conclusion, Sun is big and heavy, but far; the Moon is small and near, but light. The same factors are tuned to cancel each other. The Moon density ( a factor nearly twice greater than the solar one) explains the larger (twice) Moon tide with respect to the solar one. Somehow in a first summary, for a given spherical object and for a common mass density, the tides are ruled , by the inverse of the cube of the solid angle size.

The early theory of the tides was one of the early Newton gravity in-sights. The later tide role within a four dimensional tensor has been the greatest gift by B.Riemann to A. Einstein General Relativity Theory. These spatial tensor components in the near field are the tides we are considering. The same Riemann Tensor in the far field is associated with the foreseen (1916) and recent (2016) gravitational waves (GW) discovery.

### 1.1 May gravitational waves make any tidal extinctions?

The high energetic GW intensity and their Riemann components are very much diluted by the distance. They cannot be really relevant or dangerous to our Earth in our galaxy. However , in an imaginary galaxy where an ideal observer is located in an "Earth"-like planet, placed very near ( a few parsec distance) from a giant binary Black hole, as the largest known, in the OJ287 system ($1.8 \cdot 10^{10}$ main object and a companion of $1 \cdot 10^{8}$ solar masses, a system very far away from us), then these BH spinning and in near collapse stage, would produce so large GW that they could be dangerous for their tidal influence . The range is as large as the tidal extinctions discussed in this article. The huge inner twin BH collapse could produce (observed at parsec distance) intense amplitude space GW able to deform widely and to make in oscillation ( for a few hours period) the "Earth" radius , even by a km height tide altitude. Such an "Earth" presence and survival near a huge BH binary system is very hypothetical and quite unrealistic.

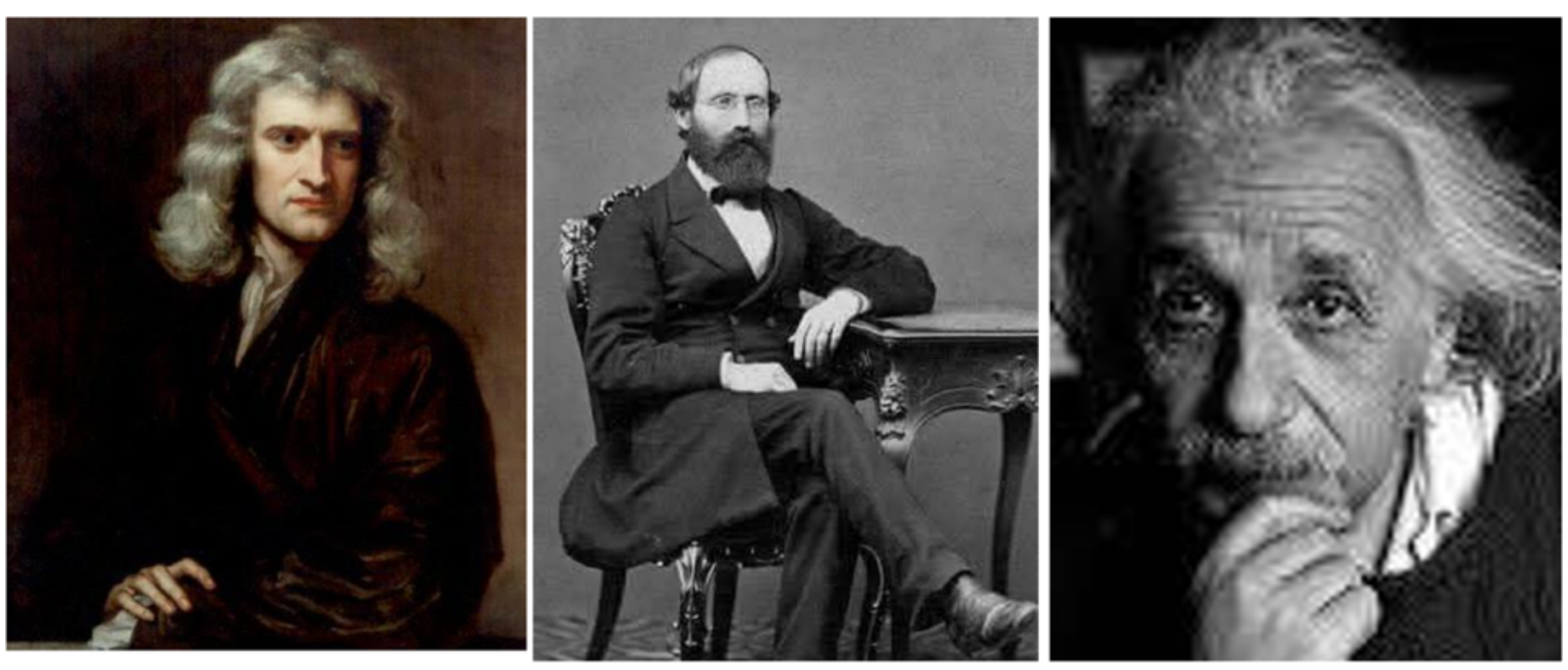

**Figure 4:** The three giant scientist of the gravity whose discovers described tides in near and in far field regimes.

## 2. Recent discovers of a dwarf planet, 2017 OF201, in far Kuiper belt

Several months ago the article [1], discovering a new additional dwarf planet , had been published. Its presence imply a population (hundreds to thousands) of such unseen dwarf planets in far elliptic orbits. This discover reinforces a previous earlier historic one [3]: The Kuiper belt geometry is made by a disk of 30 AU (Astronomical Unit $=1.5 \cdot 10^{13}$ cm. and a width of nearly 1 AU. Therefore, as an order of magnitude the whole Kuiper belt volume is $V_{Kuiper} = 1.7 \cdot 10^{44}\ cm^3$.

Based on the discovery of 2017 OF 201, researchers estimate that hundreds of similar "dwarf planet-like" or a large population of trans-Neptunian objects (TNOs) (spread in wide disk) likely exist in the outer solar system. Because 2017 OF201 is only visible for about 1% of its 25,000-year orbit, it implies a large undetected group of sources. Incidentally, these results suggest that the existence of 2017 OF201 may be difficult to reconcile with this particular instantiation of the Planet X hypothesis. This dwarf planet, 2017 OF201 , is estimated to be roughly 700 kilometers in diameter, placing it among the larger candidates in the far outer solar system.

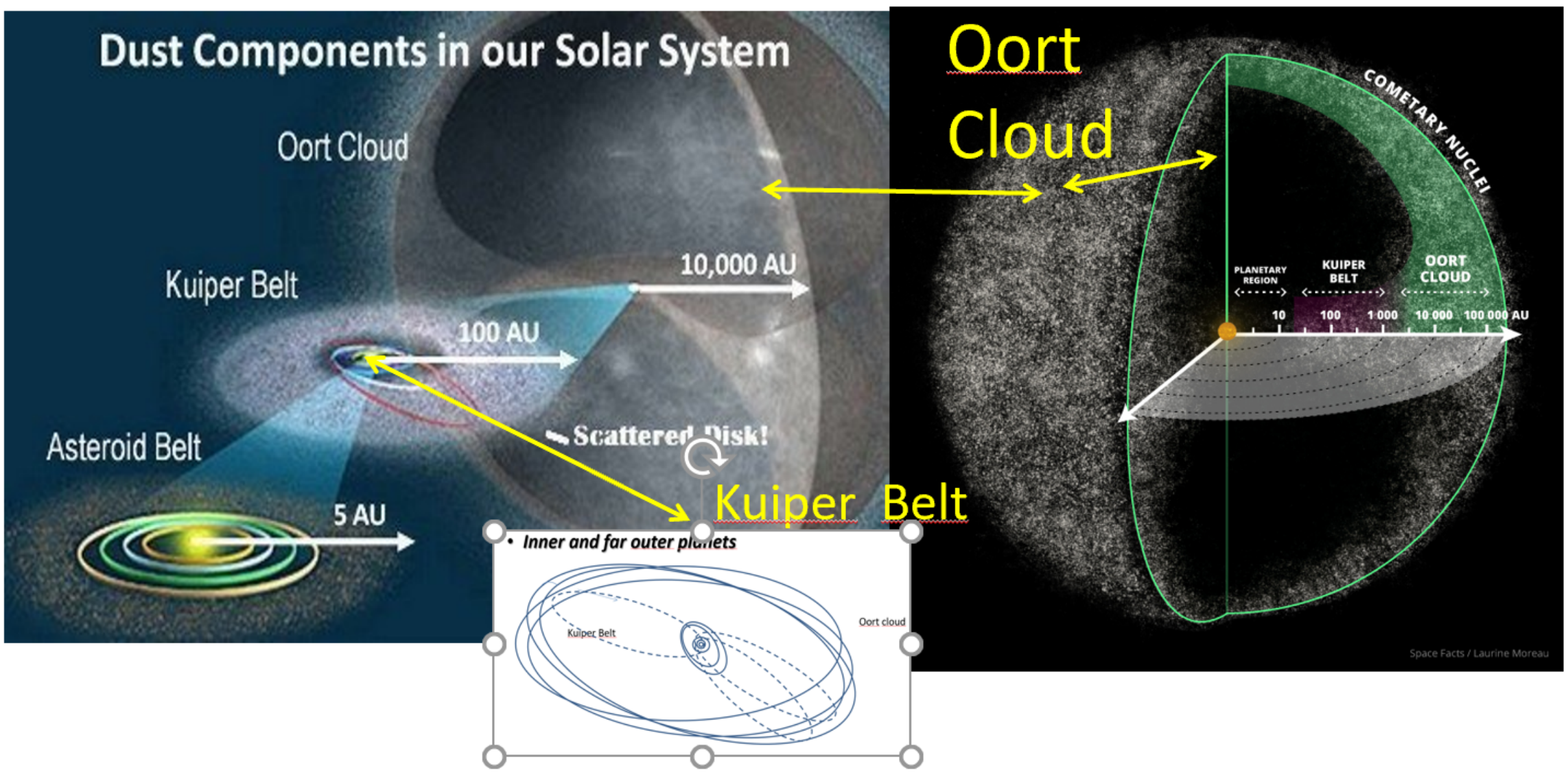

**Figure 5:** The solar system geometry where mini-planet may be confined. Because their low size and area their extreme reflected lights dilution ($d_1^{-2} \cdot d_2^{-2}$ => $..d^{-4}$), they might be hidden and hardly observed in Kuiper Belt or even more far , in Oort Cloud. Their long life orbit in far hyperbolic trajectory edges may allow their presence to be quite difficult to be discovered.

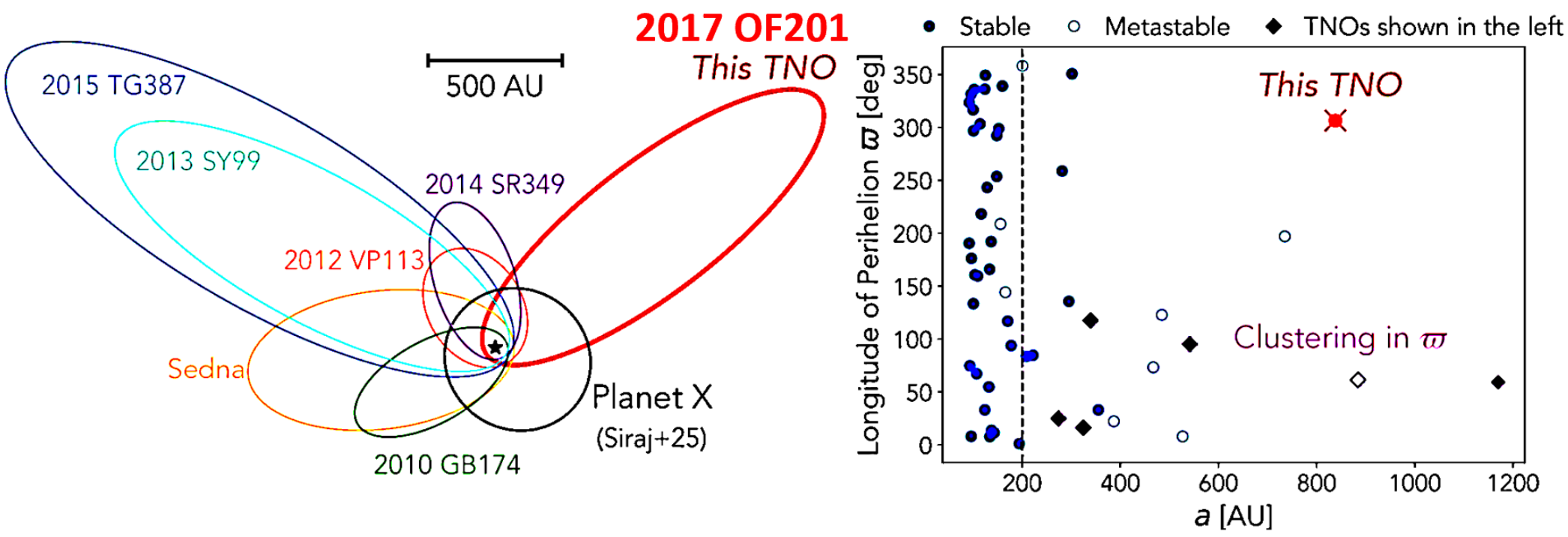

**Figure 6:** The orbit of the new , very distant dwarf planet in Kuiper Belt edges, discovered only recently [1]: 2017 OF201.

## 2.1 The probability of an encounter from Kuiper belt

The total mass (of such an undiscovered mini-planet population) could be as significant as 1% to 10% of the Earth's mass. Assuming our Moon, nearly $10^{-2}$ terrestrial mass , or a micro-moon smaller than $10^{-3}$ or even $10^{-4}$ terrestrial mass , their total number may range from a minimal 10 moon-mass up to ten thousand micro-moon-like objects. The probability P for a near encounter at

6 Earth radius is

$$P_{event} = \sigma \cdot n \cdot l \quad (3)$$

Where $\sigma$ is the cross section, assumed a spherical area of 6 Earth radius: $\sigma = 4 \cdot \pi \cdot (R_E)^2 = 5 \cdot 10^{19} \cdot cm^2$. The number density $n$ is related to the characteristic dwarf planet mass, ranging from a few, 1, (for a Moon like mass) to a $10^3$ for a micro-dwarf planet (a hundred times smaller than our Moon). This mini planet path flight along a 4 billion years at an average $20 km/s$ is $l = 2.4 \cdot 10^{23} cm$: Therefore the probability for an encounter ranges from:

$$P_{event} = \sigma \cdot n \cdot l = 0.07 - 700 \quad (4)$$

Such an approximated probability would have a mostly dwarf planet visiting us from the Kuiper Belt. The result is not precise, but it is quite auto-consistent with the current tidal extinction scenario .

Nevertheless, we should remind the reader that the Moon was , very probably, a fragment made by a full Theia hit , a dwarf planet collapsing inside a terrestrial area : just $1/(36) = 2.77\%$ of the area (around 6 Earth radius) mostly considered in our present model.

The Oort Cloud might also play a role as a reserve (by a much larger dwarf-mass deposit, up to ta housand times more than Kuiper masses). However , the Oort wider volume is so extended with respect to the Kuiper disk that the probability of Oort sources encountering Earth is more diluted and small.

The whole population discovery, using archival data, highlights that many objects in this region are currently too faint to be observed with existing technology. These findings, along with data from objects like Sedna, suggest that the outer solar system is far more populated than once thought. The Kuiper belt and possibly even the far Oort Cloud store, unobserved, several hundreds or thousands of such mini-planes.

### 2.2 Nearby fast asymmetric passage

However, a heavy dwarf planet ($m > M_{Moon}$ flyby could most probably occur at far Earth distances (for instance $d > 6 \cdot R_E$), the lighter ($m << M_{Moon}$ and smaller size mini or micro dwarf planet, whose number is much more abundant in the solar system and possibly in the Kuiper belt, might cross even more often near our planet. For example, at $d << 2 \cdot R_E$. Assuming a planet speed around $20 km/s$, the consequent tides are very rapid and stronger toward the nearest Earth side than towards the far one. This different among the two side forces produces an asymmetric tide whose damage (sea tsunami and soil deformation) could be remarkable. These smaller and abundant micro-dwarf planets (at limit of asteroid size), being at nearer altitudes, may amplify by the inverse of the fourth power of the distance their deposited tide energy. The short speed scale time of the passage may reduce the effective acceleration action and the resulting height and deposited energy .

## Planetary collisions, near and far flyby : observational traces

Collision of planet-mass objects with known planets may explain various "anomalies" in the solar planetary system . A collision of a Mars-sized object with Earth could have formed our moon

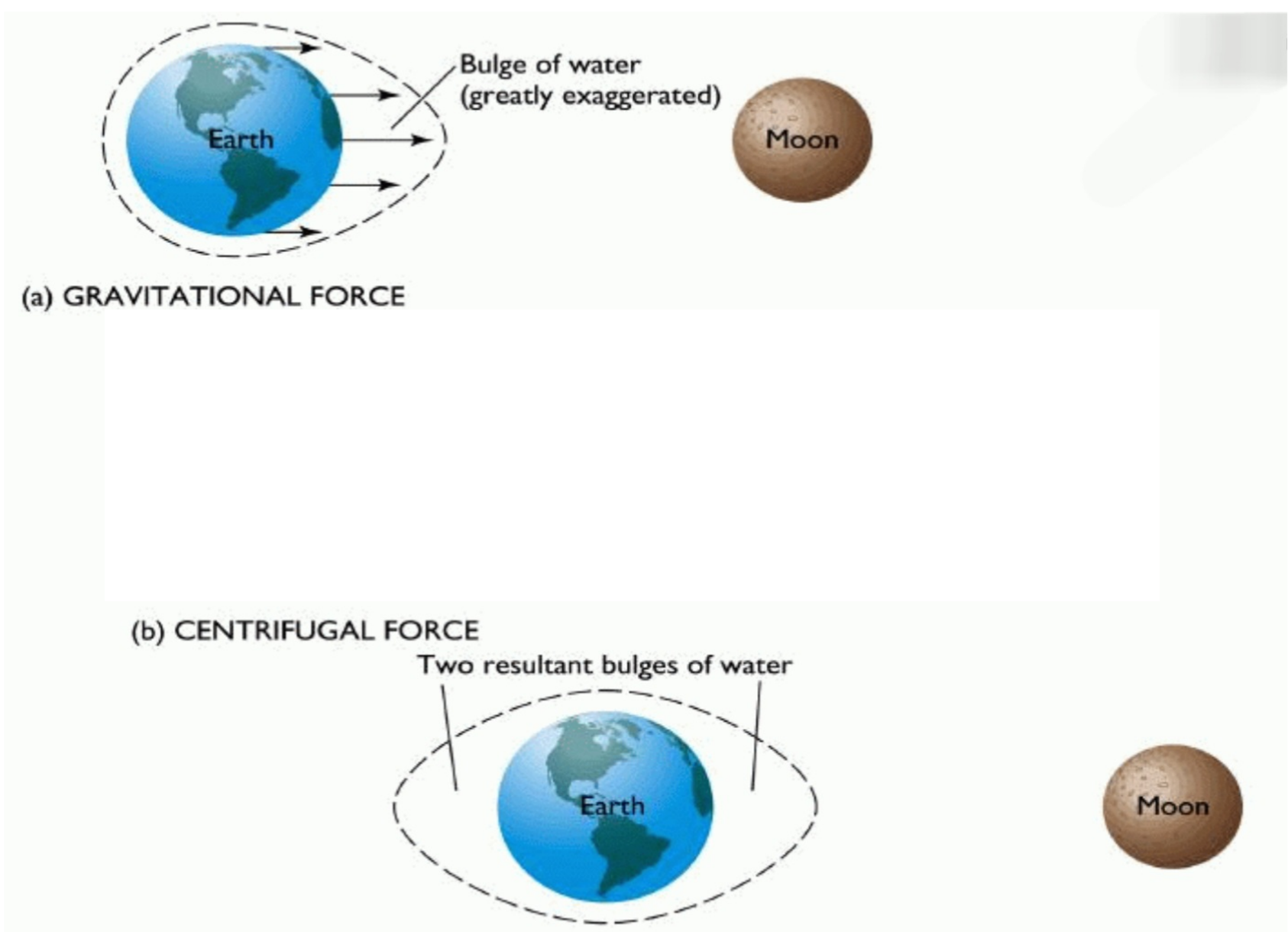

**Figure 7:** The simplest description of two extreme tide effect: the asymmetric near field (usually a fast one due to smaller bodies) in figure above, whose nearby growing tide height is proportional to the inverse of the square power of the body distances. Below the more common figure for a far field tide, at slower or nearly symmetric, in equilibrium twin forces ( as for our Moon ), a tide proportional to the inverse of the third power of the body distance from the Earth encounter

and tilted the spin plane of Earth relative to the ecliptic. Such a collision could also have formed Pluto's moon Charon, tilted Pluto's orbital plane relative to the ecliptic plane by the observed $17.1^0$ and changed it into its observed high eccentric orbit, $e = 0.253$. Otherwise recent understanding describe Pluto and Charon as candidate dwarf planets in the inner Kuiper belt regions.

The inclination of Mercury's orbit by $7.0^0$ and its high eccentricity, $e = 0.206$, could also have resulted from a near by deflection with a visiting planet. Collision, or near encounters, with visiting planets could have tilted the spin plane of the other planets/moons/satellites relative to their orbital plane (Mars: $25^0$, Jupiter: $3^0$, Saturn: $25^0$, Neptune: $28^0$, Uranus: $98^0$ and Venus: $178^0$).

Moreover, the visiting planets and their capture by the giant planets can explain: (a) why there are moons/satellites, such as Phobos and Deimos of Mars, with completely different density and composition than that of their planets, (b) why there are moons with unexplained retrograde orbits, such as Triton around Neptune, and (c) why the orbit planes of some moons, mini-moons, and asteroids are tilted relative to the equatorial plane of their planets.

### 2.3 Pro-grade and retro-grade moons in solar system

In particular, why did we find moons with retrograde orbits around the heavy planets Jupiter, Saturn, and Neptune, and those with pro-grade but very eccentric orbits, such as Nereid of Neptune, which has the most highly eccentric orbit of any known planet or satellite. Among the nearly 455 known moons (including Pluto, but not Earth), in updated NASA site on Moons , $https : //science.nasa.gov/solar - system/moons/$ there are about 314 candidates with retrograde

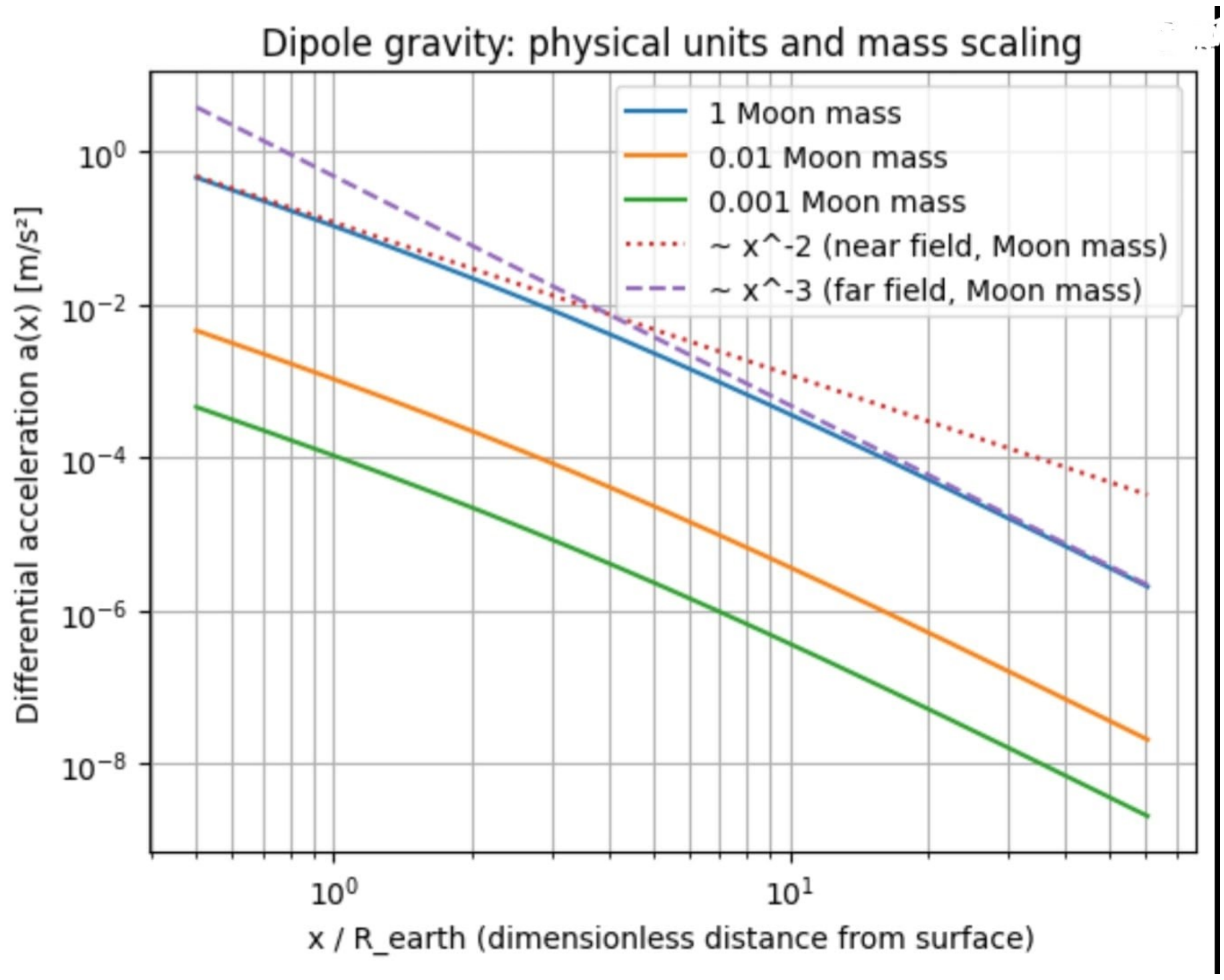

**Figure 8:** The different acceleration leading to tides, their distance dependence in nearest altitude (note, not their center of mass). Here we use a distance x (in Earth Radius unity $R_E$) , for near field or far distance regime. The tidal law for symmetric tide at far distances ($x > 10 \cdot R_E$) is nearly tangent to a inverse dashed cube power line. At a near field, the asymmetric tides are shown with a dot tangent line related to the inverse of the square distance $x$. In those near regimes one should consider also the Roche limit radius before that the tidal forces by the Earth could exceed the self-gravity holding the mini-planet together. The smaller mini-planet masses accelerations and their consequent smaller tides are described by the orange and green lines. Because of the inverse cubic law of tides with distance (dashed line, far field tangent) , the law is nearly comparable with the more exact colored curve up to 10 or even 6 Earth radius. Therefore one may apply the usual, simple, lunar approach for those distances. Only at a very near field passage one might need a more detailed exact curve , as the three colored ones above. The time duration, if fast, (when it is below the ocean time response, of about an hour), it may reduce the tide height for the sea by an additional factor proportional to the $(d/v)^{-2}$

.

orbits : 90, among 115, around Jupiter; 9 among 29, around Uranus; and 5 among 16, around Neptune; 210, among 292 around Saturn . About the same number of pro-grade moons have a large inclination ($\sim 20^0$) with respect to the equatorial plane of the planet they orbit. This is consistent with the fact that pro-grade and retrograde approaches should be equally probable, but, tidal interactions between the heavy planets and their retro-grade moons bring them closer and lead eventually to their tidal capture or disruption into rings like those around Saturn.

As mentioned , retro-grade moons are the majority, 314 , 69% versus 141, 31% of pro-grade ones. The largest moons are mostly pro-grade ones. The huge number of retrograde are small captured fragments of irregular moons. The largest number of retrograde mini moons are dominated by the Saturn Norse retrograde group.

### Recent encounters of dwarf planets hidden in terrestrial-lunar records

Most of these could have happened in the early solar system when the collision rate was much higher, as is evident from dating of moon and planetary craters . However, there are observations which suggest that near encounters of planet-like objects with the Earth-Moon system have occurred also more recently. For example, fossil corals show that the rate of decrease in the number of diurnal rings during annual cycles, i.e. the rate of decrease in the number of days in a year, has changed suddenly into a slower rate at the end Devonian (10). Since the lengthening of the day is due to slowing of the rotation of Earth by the well understood moon's tidal forces, it implies that at the end Devonian the moon-Earth distance increased suddenly by a significant fraction. Such an increase could have been induced by a tidal pull in a nearby passage of a visiting planet/moon (a crash into Earth induces a discontinuous change in the length of the day rather than a continuous one). The tidal pull increases the Moon-Earth separation if its projected trajectory in the Moon-Earth orbiting plane does not fall between them. The maximal radial kick to the moon is when the trajectory of the visiting planet is perpendicular to the Earth-Moon line on the far side of the moon. For a passing distance $d$ it is

$$V_r \approx \frac{2GM_p}{v}\left[\frac{1}{d} - \frac{1}{d+D}\right], \tag{5}$$

where $D = 380000\ km$ is the Moon-Earth distance and $v$ is the velocity of the visiting planet relative to the Earth-Moon system (if $V_p(\infty) \approx 0$ then $V_p(1\ AU) \approx \sqrt{2}V_E$, and $(\sqrt{2}-1)V_E \le v \le (\sqrt{2}+1)V_E$ with $V_E \approx 30\ km\ s^{-1}$). For $d \gg D$, this tidal velocity is quite small. However, if the visiting planet passes close to the moon in a direct orbit at a distance $d \le D$, then $V_r \sim 2GM_p/vd$. Such a tidal velocity can significantly increase the moon's distance from Earth and divert it into an eccentric and inclined orbit (as observed). The eccentricity will be damped by terrestrial tidal forces. For $d < 0.1(M_p/M_E)D$, a planet can even eject the moon from its Earth-bound orbit. The passage of a nearby object can also slightly change the Earth orbit around the sun and its spin orientation. Even slight changes in the orbit or spin orientation of Earth can have dramatic effects on climate, sea level, and glaciers. The few observed radical changes of past terrestrial spin , slow down rate, can be caused by such a dwarf planet encounter and its deflecting lunar-Earth orbits. These traces seem sometime to be correlated with a few mass extinctions. For example, The largest mass extinction event in Earth's history, the "Great Dying," along the Permian Extinction (approximately 250 Million years ago) occurred during another period of rapid, dramatic rotation slowdown.

## 3. The tide height by lunar presence as an example

In the very early Earth epochs, after the Theia hit and the Moon birth, our natural satellite was much closer to us, The Moon was, in the earliest billion years, much closer to Earth, and its distance was expanding from us quite faster than now by stronger tides. They were (and are ) the cause for such distance growth. For the same reason, the Moon was forced in billion years to face almost always toward our Earth. In very far future stages, billion years from now, the Moon distance will increase so much that its tides will decrease much more than solar ones. Then the solar-tidal role (nearly unchanged) might even be comparable or might exceed the lunar one.

In the case of oceans versus our solid Earth crust, the height of the water increases only (on average) half of the above result, or just

$$\Delta h = 0.27\ m \tag{6}$$

Anyway we shall often use the previous half a meter size, in next tidal flyby models for the Earthquake tides. We shall also discuss the three extreme regimes for a more general formula:

The above, simplest, equilibrium tidal force regime, where the tides have time to grow as above, The height grows as the cube of the inverse of the earth-Moon distance d: $h = ..d^{-3}$. .

The impulsive regime is ruled by a very fast planet speed (for a given flyby passage), where the short tidal time does not allow for a full height growth. This requires a more dominant quadratic speed suppression term in the height tide formula d: $h = ..(d^{-3}/v^2)$.

The nearby and fast tide event where the gravity forces on both the Earth sides are not equal. The nearer is much more dominant than the other far side. Therefore, the growth of tide forces depends mainly on the inverse of the square distance $h = ..(d^{-2})$ from the closest side. Being possibly also, usually, too fast for such a tide increase, the force, as in the previous case, might require in addition a quadratic distance decay speed : a suppression term in the height formula as follows: $h = ..(d^{-2}/v^2)$.

There are additional minor dispersion terms to be taken into account for the evolution of a more realistic tide model on geological terrestrial model. We shall introduce here only the main steps to the main tide model from the simplest ideal lunar one with few summary toward a more realistic scenarios.

### 3.1 The Energy released by a dwarf-lunar mass passage

In a first approximation the gravitational energy could be obtained for a Moon mass $M_L$ tide may be estimated by the amount of Earth mass fraction $\Delta M_E$, moved to the height increase $\Delta h$ at the tide peak, effect:

$$\Delta E = \Delta M_E \cdot g \cdot \Delta h = = M_E \cdot g \cdot \left[\frac{(\Delta h)^2}{R_E}\right] \tag{7}$$

This result may be written as:

$$\Delta E = 9/4 \left[\frac{(G M_L{}^2)}{R_E}\right] (d/R_E)^{-6} \tag{8}$$

The first square term is equal $5.6 \cdot 10^{35}$ erg, while the average Moon-Earth distance $d$ is $d = 60.27 R_E$. These values in above expression lead to: $\Delta E = 1.16 \cdot 10^{25}$ erg. A more accurate (but comparable) result could be derived considering the total terrestrial sea water mass displaced, $\Delta M_w$ , based on the the sea areas $A = 2.5 \cdot 10^{20} cm^2$. In that case only half a mass is to be considered. and the result becomes (with a visiting mini-planet mass $M_v$ ) and its Moon-Earth distance d in Earth radius unities is :

$$\Delta E_L = 6.6 \cdot 10^{24} \cdot erg \cdot \left[\frac{(M_v{}^2)}{(M_L{}^2)}\right] \cdot [d/(60.27 \cdot R_E)]^{-6} \tag{9}$$

This energy $\Delta E$ , in case of a Theia-like visiting mini-planet passage at a speed $v = 20 \cdot km/s$, assuming a factor ten times greater mass than Moon one, for a present Moon distance, is only a hundred times larger: $\Delta E = 6.6 \cdot 10^{26} \cdot erg$ . But the probability P for such a skimming event ( respect to the rare full hit, a ratio associated to the square of the terrestrial $R_E$) assuming a visit at nearest Moon distance, $d = 60.27 \cdot R_E$, is very large . Therefore the probability of nearby visit is nearly, P= $(60, 27)^2 = 3632$ much larger than any Theia probability to reach in a central impact. For a heavier Earth-like visiting planet $\left[\frac{(M_E{}^2)}{(M_L{}^2)}\right] = (81.3)^2 = 6609.7$, at Moon distance, the tide height becomes larger: $\Delta h = 21.95$ m. The energy release will be nearly four order of magnitude larger. Indeed the corresponding deposited energy becomes more significant: $\Delta E_T = \left[\frac{(M_E{}^2)}{(M_L{}^2)}\right]^2 = 4.36 \cdot 10^{28}$ erg. The much nearer approach, the very power-full sixth exponent term $(d/R_E)^{-6}$ in the tide energy expression, lead to output comparable or even superior to the most disruptive energies associated to an eventual , large diameter Asteroid impact (10 − 20 km) at realistic speed $v_A = 20$ km/s , an order of magnitude:

$$\Delta E_{Asteroid} = 10^{30} \cdot erg.$$

. For instance a Moon, or a mini planet Theia (ten times heavier than our Moon), or an Earth-like visiting planet approaching 10 times closer than our Moon ( anyway as near as , 6 terrestrial radius distance , or by a probability 36 times larger to such face hit) would lead to very disruptive energies anyway,

$$\Delta E_{Moon} = 6.6 \cdot 10^{30} erg \tag{10}$$

$$\Delta E_{Theia} = 6.6 \cdot 10^{32} erg \tag{11}$$

$$\Delta E_{Earth} = 4.36 \cdot 10^{34} erg \tag{12}$$

These energies and their later thermalization, are better compared with the solar energy absorbed by the surface Earth during a characteristic duration time $\Delta t = 4000$ second , for a passage at 6 Earth radius with a speed flight of $20 km/s$.

$$\Delta E_{Sun_{Luminosity}} = 7 \cdot 10^{27} ergs$$

In a week time the Earth absorbed much more solar energy:

$$\Delta E_{Sun_{Luminosity}} = 1.05 \cdot 10^{30} ergs$$

Therefore the large Moon passage will require nearly a week time to dissipate in heat (but also in soil elastic deformations) the event. A Theia tidal visit at 6 Earth radius distance, energy would require a full year to dissipate its energy. A planet like the Earth passage, would require ( in principle) a century to re-radiate the tidal energies. It is well possible that an increase of the terrestrial temperature by a factor 10% would reduce (because of the Stephan Boltzmann fourth power law on temperature) the time of dissipation by 40%. The heat created by a planet tidal passage is much less than the one by the same object full impact,. Nevertheless such tidal energy could be comparable to a much smaller size (asteroid like) full hit .

The impact encounter energy , $\Delta E_{Asteroid} = 10^{30} \cdot erg.$ in the historic Chicxulub extinction event related to the last Cretaceous-Tertiary (K/T) boundary, 64 millions years ago has the same value. The tidal tides are not leaving any craters but just costal tsunami , sea retire and explosive volcanic events . All peculiar mass extinctions signatures with no scape-goat trace.

Let us consider more general scenarios: for instance, a half Earth planet mass or a smaller mini-planet passage at a nearby distance (let assume 10 times the Earth radius one) , whose probability to occur is, nevertheless, a 100 time more probable than a central hit. Otherwise , a 10 times near distance than present Moon , (or 6 times the Earth radius), at a probability to flyby , 36 time larger than a frontal impact.

In solar system the moons or mini moons are much more abundant than planets. Therefore we consider both large planets as half an Earth, or Theia or more abundant smaller Moons, as candidate visiting mini-planets, versus a much smaller mini moons at a very near passage distances ( a few Earth radius).

### Planet and the solar accretion of dwarf planets

A reliable estimate of the masses and the flux of the visiting planets/planetesimals is not possible yet. However, it is tempting to estimate them from the assumption that the unaccounted energy source of Jupiter and its tilted spin plane relative to its orbital plane are both due to accretion of visiting planets/moons. An accretion rate of $\dot{M} \approx 5.8 \times 10^{-12} M_J\ y^{-1}$, in addition to its absorption of 55% of the incident sun light, can explain its effective $134K$ surface temperature. It implies that Jupiter has accreted $\approx$ 2.65% of its mass during the $t_\odot \approx 4.57\ Gy$ lifetime of the solar system. Jupiter's effective capture cross section is, $\sigma_J \approx \pi R_J^2(1 + M_J D_J / M_\odot R_J)$, where we assumed that visiting planets arrive near Jupiter with a free fall velocity in the sun's gravitational field, $V_p \approx \sqrt{2GM_\odot/D_J}$ with $D_J = 5.2AU$ being Jupiter's distance from the sun. The impacts are from random directions in the orbital plane. If the visiting planets have a mean mass $M_p$ then the number of random impacts are $N_J \approx 0.0265 M_J/M_p$. Each random impact deposits on the average $\sim (2/3)^{3/2} M_p V_J R_J$ of angular momentum perpendicular to Jupiter's spin, $S \approx I_J \Omega_J \approx (2/5) M_J R_J^2 \Omega_J$, where $I_J$ and $\Omega_J$ are its moment of inertia and angular rotational velocity, respectively, and $V_J$ is the planet's impact velocity. Thus, the random deposition of orbital angular momentum in Jupiter during $t_\odot$, could have tilted its spin by a mean angle

$$sin\theta_J \approx \sqrt{N_J} M_p \frac{\left(\frac{2}{3}\right)^{3/2} R_J \left[\frac{2GM_J}{R_J} + \frac{2GM_\odot}{D_J}\right]^{1/2}}{I_J \Omega_J}. \quad (13)$$

From the observed $3.13^0$ tilt of Jupiter's spin and the accreted mass, we may infer that $M_p \approx 0.5 M_E$ and $N_J \approx 16$. Similar estimates for other planets, although yielding the correct order of magnitude for the tilts of their spins, are less reliable because the inferred number of accreted planets is too small. We note, however that the past dwarf planet population, four billion years ago, could be more heavier than present one, which is mostly dominated by lighter Moon-like dwarf planet mass.

## 4. Geological imprint of tidal mass extinctions

Geological records indicate that the exponential diversification of marine and continental life on Earth in the past 600 My was interrupted by many mass extinctions [4]. The records also indicate that the major mass extinctions were correlated in time with large meteoritic impacts, gigantic volcanic eruptions, sea regressions and drastic changes in global climate [5] . Some of these catastrophes coincided in time. The reason for that is not clear. Meteoritic impacts alone, volcanic eruptions alone or sea regressions alone could not have caused all the major mass extinctions: A large meteoritic impact was invoked in order to explain the iridium anomaly and the mass extinction which killed the dinosaurs and claimed 47% of existing genera at the Cretaceous-Tertiary (K/T) boundary 64 $My$ ago. But, neither an iridium anomaly, nor a large meteoritic crater have been dated back to the Permian/Triassic (P/T) mass extinction, 251 $My$ ago, which was the largest known extinction in the history of life, where global species extinction ranged between 80% to 95%. The gigantic Deccan basalts flood in India that occurred around the K/T boundary and the gigantic Siberian basalts flood that occurred around the P/T boundary have ejected approximately $2 \times 10^6\ km^3$ of lava . They were more than a thousand time larger than any other known eruption on Earth, making it unlikely that the other major mass extinctions, which are of a similar magnitude, were produced by volcanic eruptions. However, although there is no one-to-one correspondence between mass extinctions, large volcanic eruptions, large meteoritic impacts, and global environmental and climatic changes, there are clear time correlations between them whose origin is not clear. Here we suggest that tidal effects of planetary-mass objects which pass near Earth could have caused all of these and can explain the complex geological records of the major biological mass extinctions. In fact, frictional heating of planetary interiors by gravitational tidal forces can lead to strong volcanic activity, as seen, for instance, on Jupiter's moon Io, the most volcanically active object known in the solar system. But, the moon, the sun and the known planets are too far away to induce volcanic eruptions on Earth. However, recent observations considered above and theoretical considerations [6] suggest that thousands of planet-mass objects may be present in a distant ring in the solar system between the Kuiper belt and the Oort cloud . Gravitational perturbations may occasionally change their parking orbits and bring them into the inner solar system. Various "anomalies" in the solar planetary system can be explained by collisions, capture or scattering of such objects by the planets. Passage of such objects near Earth could have generated on it gigantic tidal waves, large volcanic eruptions and drastic changes in global climate and sea level. Moreover, upon crossing the asteroids and Kuiper belts these planetary mass objects could have diverted large meteorites and asteroids into a collision course with Earth. Thus visits of distant planets may be the common cause for the various catastrophes that were jointly responsible for the major biological mass extinctions.

## 5. Tsunami and Volcanism

The tides may excite ocean and sea growth into tsunami as well as deform the terrestrial soil leading to huge Vulcan activity. This occur in spectacular way on Jupiter and Saturn Moons as observed in last decades.

### 5.1 Deccan Traps in India

The Deccan Traps are a large igneous province of west-central India (17–24°N, 73–74°E). They are one of the largest volcanic features on Earth, with lava erupted from multiple volcanic centres. The Traps consist of many layers of solidified flood basalt that together are more than about 2 kilometres (1.2 mi) thick, cover an area of about 500,000 square kilometers (200,000 sq mi), and have a volume of about 1,000,000 cubic kilometres. Originally, the Deccan Traps may have covered about 1,500,000 square kilometres with a correspondingly larger original volume. This volume overlies the Archean age Indian Shield, which is likely the lithology the province passed through during eruption. The province is commonly divided into four sub-provinces: the main Deccan, the Malwa Plateau, the Mandla Lobe, and the Saurashtran Plateau. The eruptions occurred over a 600-800,000 year time period between around 66.3 to 65.6 million years ago, spanning the Cretaceous–Paleogene boundary. While some authors initially suggested the eruptions were a major cause of the Cretaceous–Paleogene mass extinction event at roughly the same time, this theory has been rejected as a result of research into the Chicxulub impact, now thought to be the primary cause of the extinction. While some scholars continue to argue the eruptions may have contributed, it is now generally accepted that the Deccan eruptions played a minor role at most or may have even partially negated the effects of the impact. The Deccan Traps are thought to have been produced in major part by the still active Réunion hotspot, responsible for the creation of the modern Mascarene Islands in the Indian Ocean.

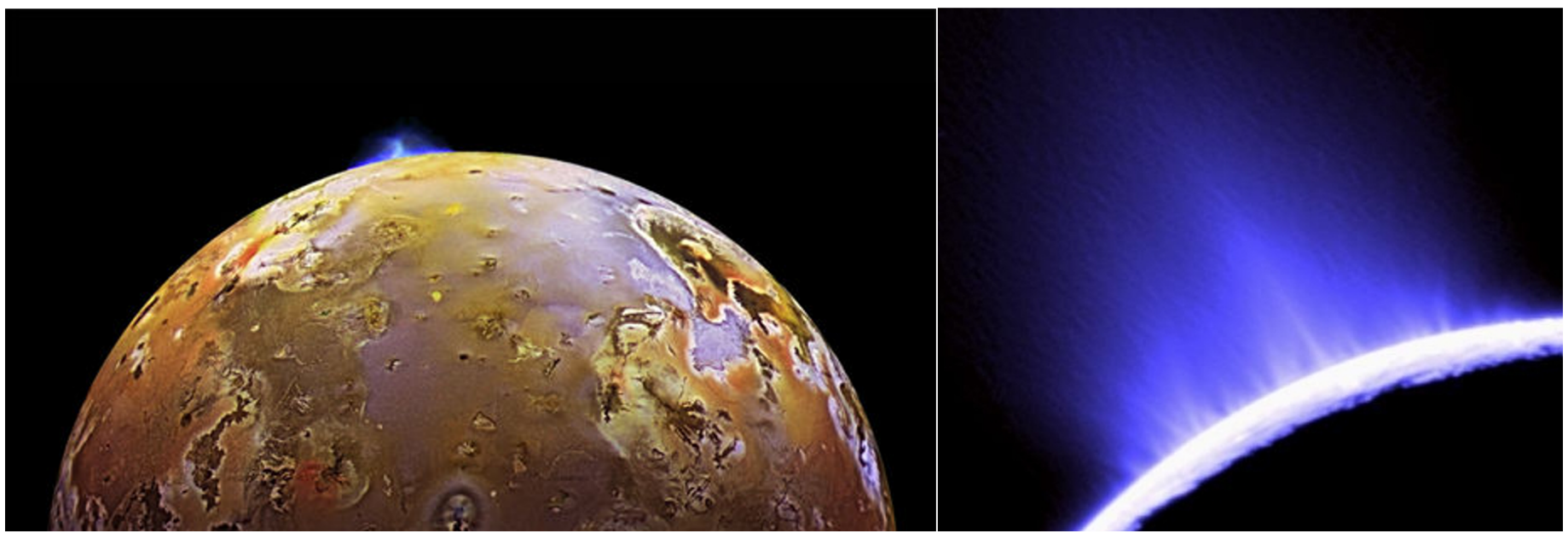

**Figure 9:** The Io moon volcanism expelling hundred kilometer gas (left image) and the similar tidal emission made by Saturn on Enceladus, expelling water jets from the icy moon of the satellite.

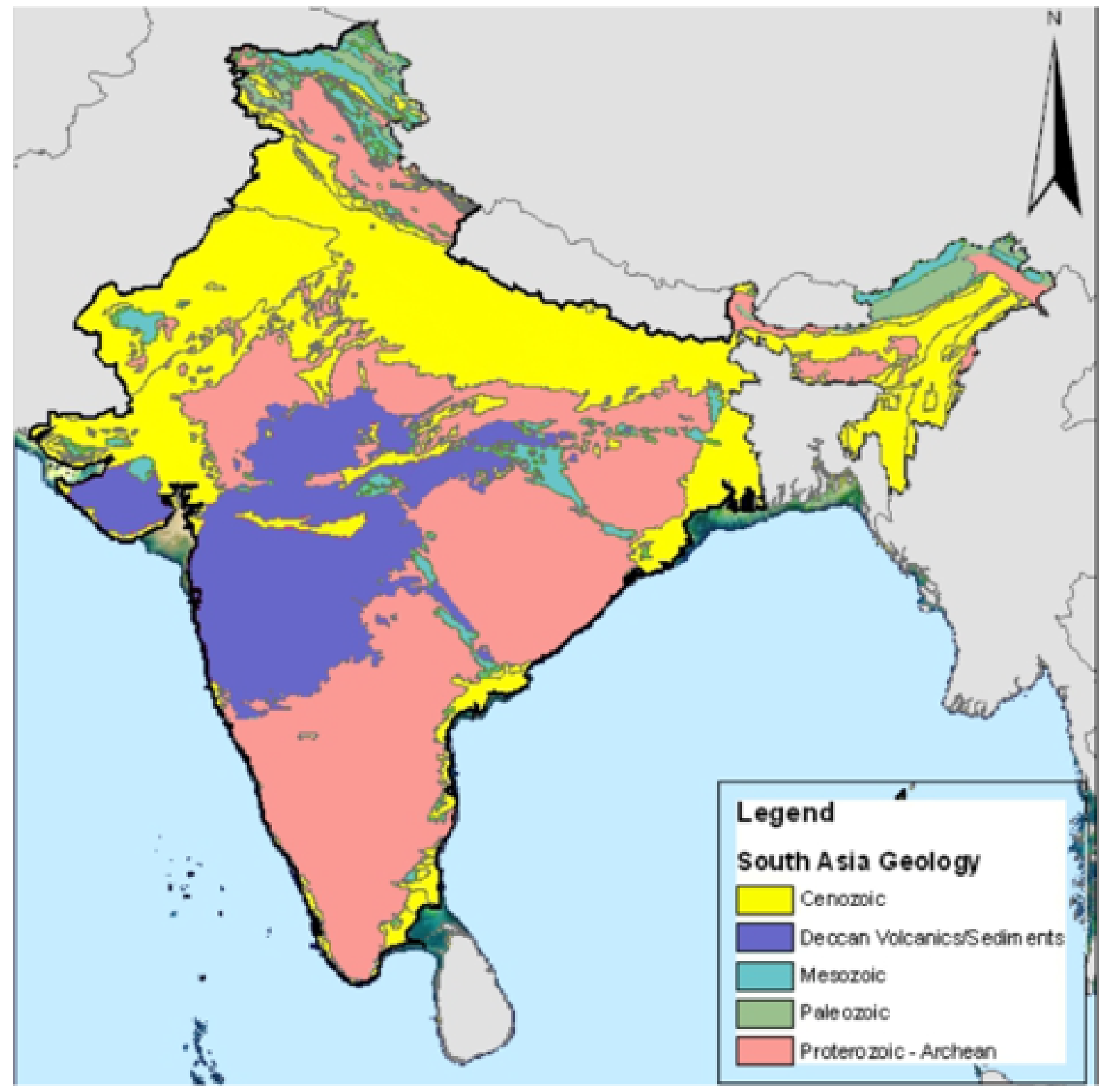

**Figure 10:** The India surface by huge Tsunami and vulcanize event 64-66 million years ago

Let us remind , as an example, the huge terrestrial tsunami, occurred on on December 26th, 2004, when a massive 9.1 magnitude earthquake struck off the west coast of northern Sumatra, Indonesia. The third largest earthquake ever recorded lifted the sea floor several meters, causing tsunami waves to ripple out in all directions and race across the ocean. The waves reached just 15 to 20 minutes after the earthquake. The waves in some locations along the Sumatra coast were reported to be 30 meters high. The energy released was comparable to a million and a half Hiroshima nuclear explosion. A negligible energy release (a part over ten thousand) respect the tidal passage of a Moon-like dwarf planet ten time nearer than present lunar distance. But in principle an earthquake by a huge mass acceleration capable to deflect the near field gravitational metric tensor in an a detectable edge, for present or future gravitational wave array [7].

## 6. The animal survival in mass extinctions

There is a very peculiar imprint of the tidal mass extinction. The contemporaneous extinction of both ground and sea anima. As we know from known tides the retire of the sea leave sea animal at air, leading in a few minutes to their death. On the other side, the tsunami wave invasion of the coastal soil would kill all the air leaving life. Only amphibious animal might survive. As well as long time flying birds. These are, mostly , the main survive trace left in largest mass extinctions.

### 6.1 Life survival from the sea tides and the shoaling amplification

The most of the sea tides energy is first stored in their gravity waves altitude. Than in their kinetic run toward the coast soil. Finally the height of the tsunami wave-front is much increased while encountering the coast soil morphology, leading to a power-full amplification. It is called the Shoaling effect. The shoaling effect can amplify wave heights by a factor (3-4) of several, depending on bathymetry and coastal geometry.

For instance, a tsunami by a Moon at ten times nearer distance (6 terrestrial radius) would experience not just a $0.27 \cdot 10^3 = 270$ meter altitude, but a few kilometer final tsunami height on front of the coast. Such a huge water wave front would reach and destroy world coastal boundary and areas, leaving sudden huge empty sea side areas of water beyond. Both the aquatic and the coastal life will suffer a sudden death. The changes in air atmosphere by volcano pollution and the temperature changes could explain the later death of many more species.

## 7. Tidal Extinction summary: An event sequence in few hours

Tidal tsunami may derive by a Moon-like dwarf planet passage near the Earth. The very recent dwarf planet discovered [1] in the far edges of the Kuiper belt could be the example. At a distance from Earth ten times nearer than our Moon, the passage of a Moon-like dwarf planet may excite oceanic tidal waves, more than 1 km high. They can flood vast areas of continental land and devastate sea life and land life near continental coasts. Blowing a correlated volcano activity. The spread of ocean waters by the giant tidal wave over vast areas of land and near the polar caps might later on enhance glaciation and sea regression.

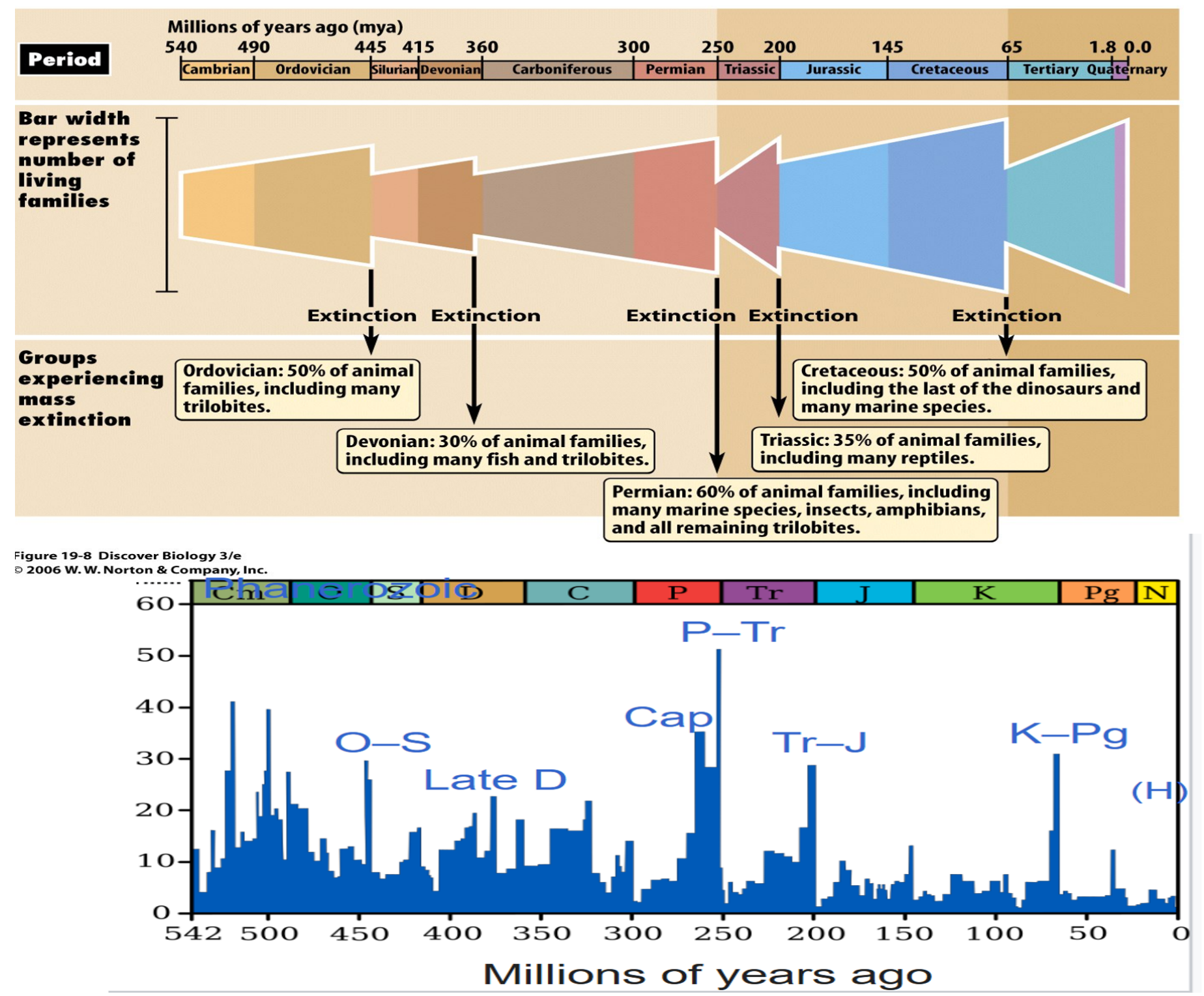

**Figure 11:** The animal mass extinction in their survival ratio, in last half billion years.

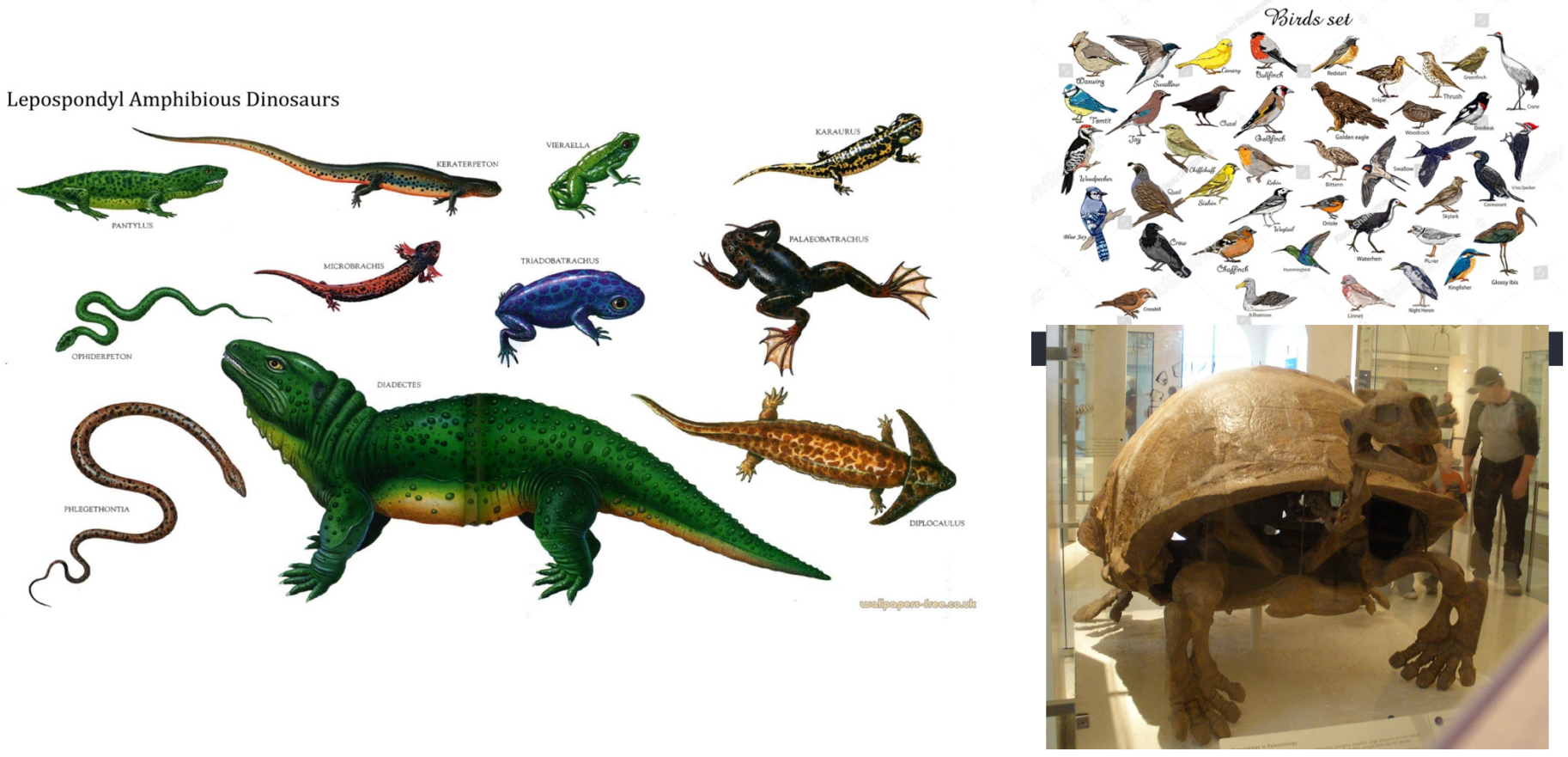

**Figure 12:** The animal that survived the main mass extinction were mostly amphibious and birds. The tsunami waves by tidal extinctions, would produce survival conditions preferentially to those animals.

An intuitive estimate of the tidal phenomenon is an ideal narrative of a Moon-like near encounter, in the past (or future) geological epochs, by such a dwarf visiting planet. Let assume, for simplicity, that these dwarf planet, just for simplicity as large as our Moon, will reach a distance ten time nearer than our present Moon. Its coming at a characteristic (hypothetic) speed of $20 \cdot km/s$

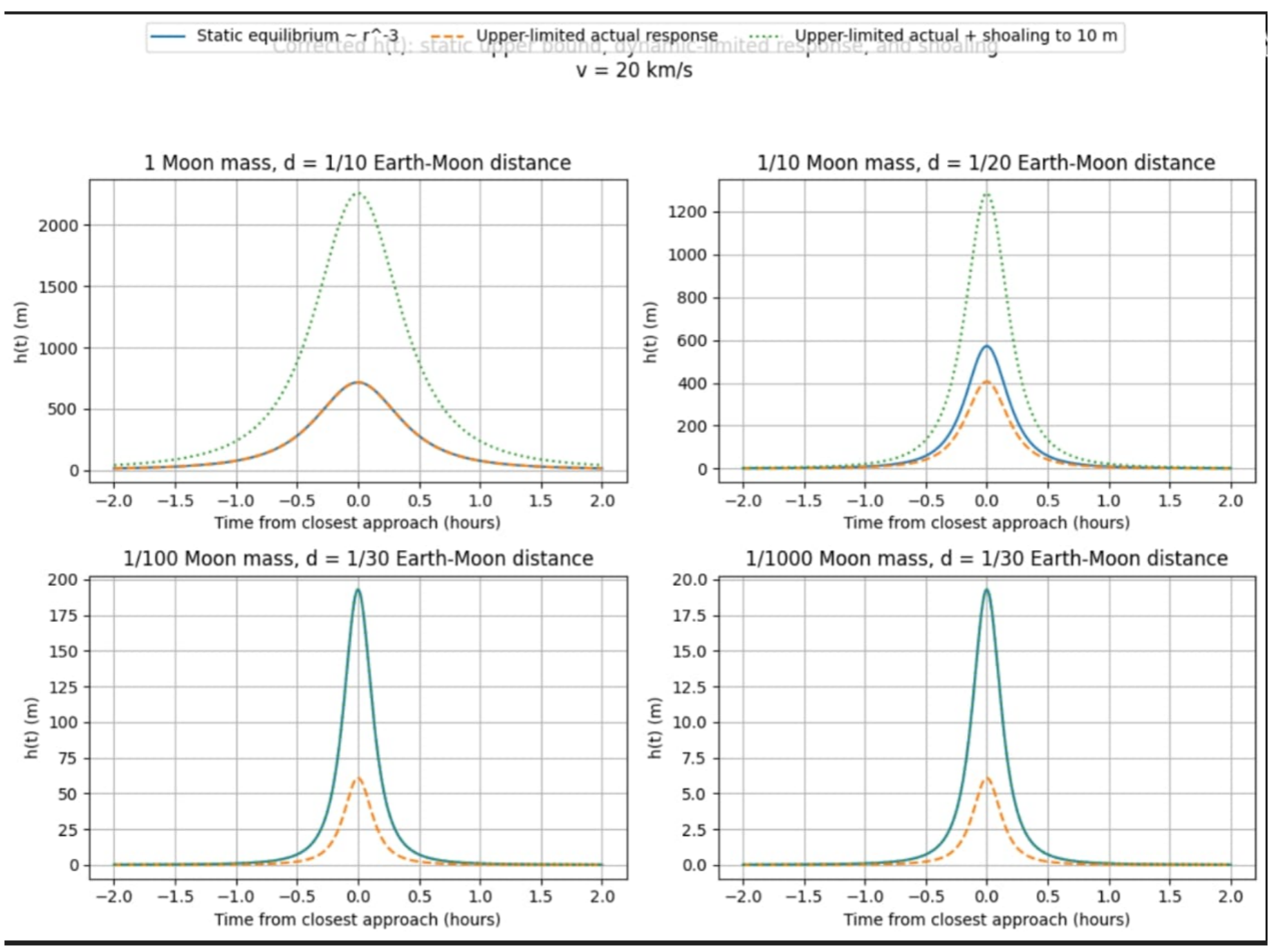

**Figure 13:** The nearby encounter made by a lunar moon mass as a mini-planet, at different distances, all at v= 20 $km/s$ speed. The equilibrium case, at ten times nearer distance than our Moon, top left side, lead to a thousand times greater tide as shown in orange, while the higher green dot curve stand for the coast wave increase in the hit of the ocean-soil boundary (the shoaling effect). On the top right side, the smaller 1/10 mass Moon-like passage , at a nearer distance , that compensates the smaller mass. The ideal tide is the continuous blue curve, the amplified shoaling wave is again in dot green curve, while the real (partially suppressed wave by the fast near arrival), is the lower dashed orange curve. The lower twin figures, left (1/100) moon mass and (1/1000) moon mass, are showing the much nearer ideal (blue) and real (dashed orange) suppressed tides by fast encounters

would be, by the eyes of any hypothetical observer, mostly poorly noted for years. Only a couple of months before its main arrival, it will start to shine visible at night, as a little planet as luminous as our Mercury one. Only at very late times, five and a half hours before the main encounter, the presence of such visiting dwarf planet would be enlarged and it will be show itself dominant in the sky by an angular size diameter $0.5^o$, comparable to our present Moon. A twin moon presence in our sky, while the dwarf visiting planet is moving and growing in its size. Within a few hours, its angular size will enlarge by a factor ten, as much as $5^0$ in diameter size. Meanwhile huge tsunami should rise , from half up to a few a kilometer height, with huge Earthquake and Vulcan eruptions. . Meanwhile the visiting dwarf planet reached its nearest distance from Earth (40.000 km) and its trajectory will be gently bent in an arc. This dwarf planet will soon fade away as it arrived sequence, leaving a huge chain of energy injections, dispersions and death on Earth. Because of the modest factor "ten" in the solid angle presence, any superficial observer, could first foresee only an under-estimated energy release, just by a factor ten respect our Moon tides (in a half a day, of 27 cm.). The dwarf-planet short duration passage, would be of nearly an hour versus our lunar tide of half a day.. Therefore it might also increase by its rapidity to a higher power respect slower Moon tides. By an additional factor ten. These two small increase toward a factor hundred, as we had shown, is totally misled. The tide gravity height, we have reminded , grows with the inverse of the cubic of the object distance. The tidal energy is associated to the square of that height, leading to its intensity growth as the inverse of the sixth power of the distance. Incidentally as the object apparent angle aperture increase by a factor ten on the sky, the tide energy injection a million times larger the quite and slow Moon tide energy , at usual far (60 Earth radius) . A near passage tide amplification by nearly seven order in power burst, whose final signature, shown in figures below, is ending in a final heat dissipation: An energy released in an hour window time, comparable to a week of the integral solar luminosity on Earth. One may also compare these energies to the Hiroshima nuclear bombs. The total amount of tidal energy deposited (in this Moon, ten times nearer visit to us) is comparable to ten billion of Hiroshima bombs exploding within 4000 seconds. Finally we remind that the approximated world wide nuclear power reserve is estimated to be about 12.241 nukes. Each of them an average, about 200 kT (kilotons of Nitroglycerin), (15 times the Hiroshima bomb energy). Therefore a total nuclear energy by nukes on Earth , around $10^{26}$ erg, is nearly 60.000 times smaller than what we estimated by a tidal energy release . The tsunami wave power is mostly dissipated in horizontal waves by several very fast tsunami (and Earthquakes) on the sea cost front. This lethal hit is facing all continent boundary at sea-soil edges and their inner coast soil. Sometimes, possibly a companion asteroid (of the vistiting planet) may impact. These models may fit better with some puzzle of the past mass-extinction on Earth.

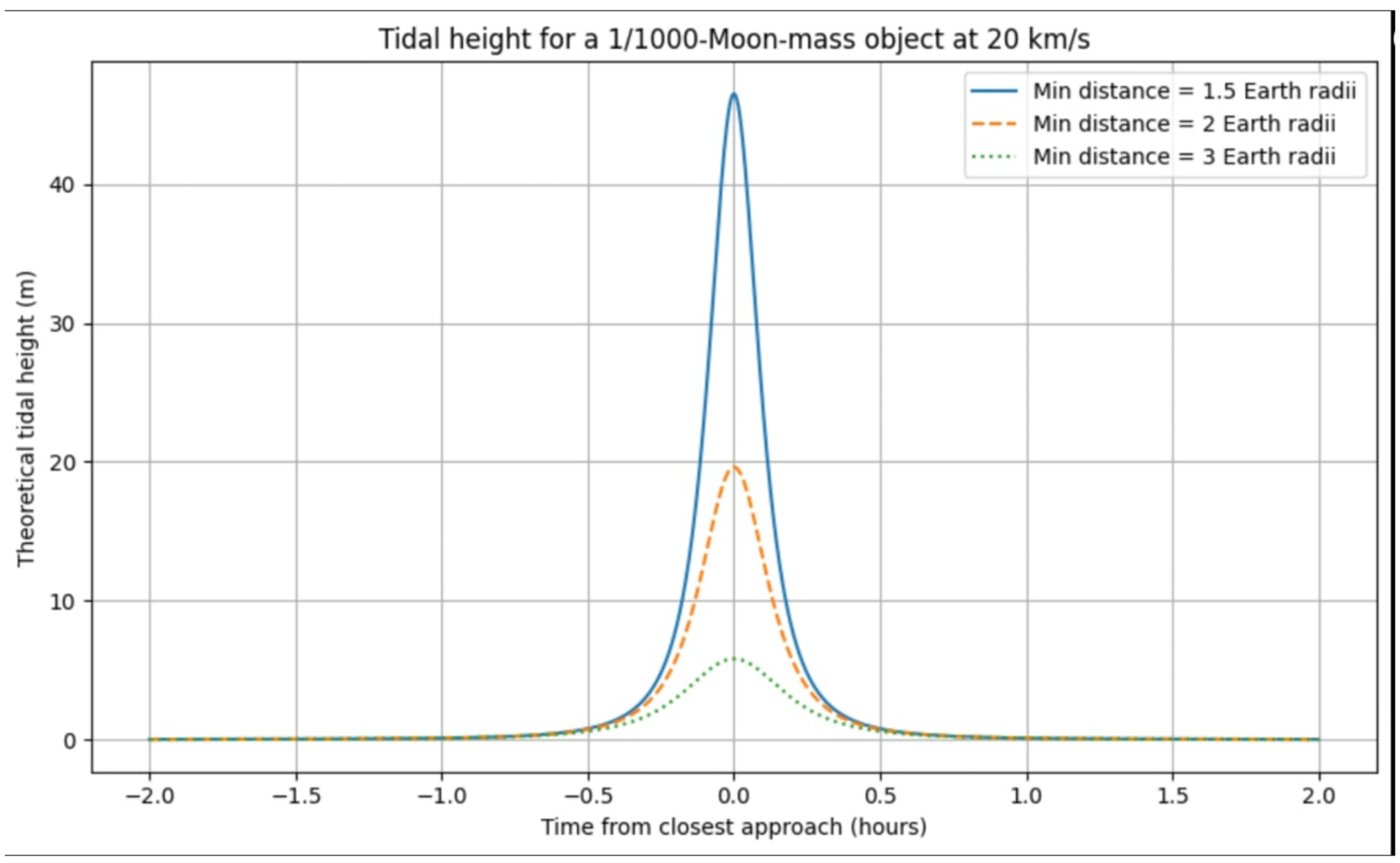

**Figure 14:** The nearby encounter made by lightest mini-moon like planet flyby, a thousand times smaller than our Moon, at nearer different distances. The Roche radius is ignored here, assuming that the mini-planet mass density (rock) is able to survive the tidal disruption. The distance in present description is considered as the nearest distance of the center of mass of the mini-planet from the surface (not the center) of our Earth, while skimming in flight. The wave altitude are not as large as to other examples for tidal mass extinction. But they would release in principle a huge localized energy along their tangent trajectory with asymmetric clustered Earth disruptions and mass extinctions

The consequent lessons for us, humans, to overcome such an eventual extinctions are: deep sky inspection to seek weakest far dwarf planet sources, once found, alert the incoming events as soon as possible. in case of an asteroid size visit, try to divert it [8]. A prevention might be also to organize, on the top of mountains chains ( at 2-3 km altitude), secure refuge for human and life for surviving such (rare but not, impossible) events. The tidal mass extinction, its rate and form, may also explain the famous Fermi Paradox: Where is everybody, or , why they are not yet here? The answer could be, thatlife is unstable and short. Evoluted civilization are more and more fragile. Social hate, nation ideological fights and consequent violent wars could self-destruct our existence. Astrophysical tidal extinctions could more rarely cancel any advanced civilization, making a life reset toward a lower, primordial, silent level.

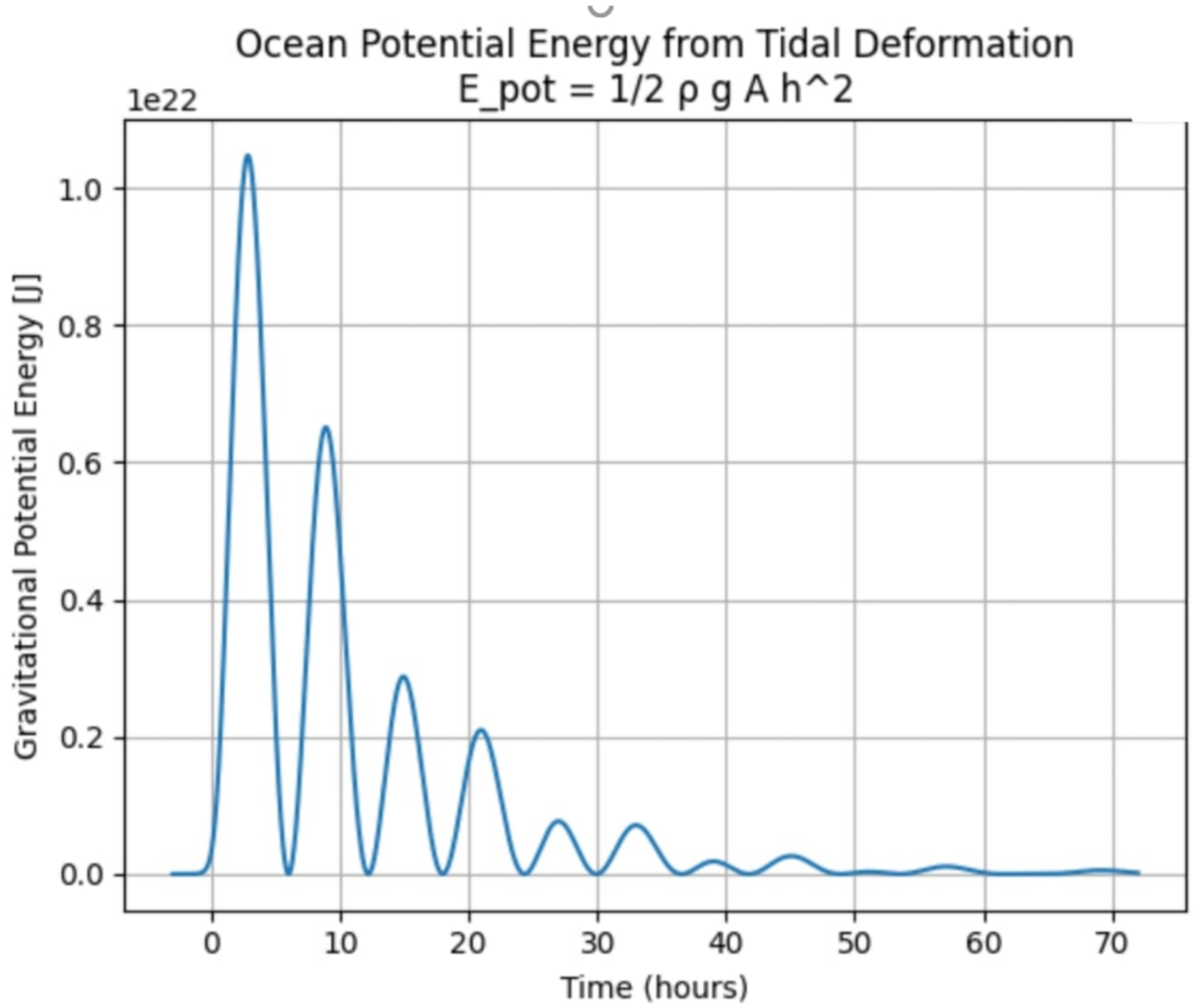

**Figure 15:** The energy dissipation with a characteristic damping and dissipation by heat radiation after the tidal disruption, assuming a Moon-like mini-planet passage at 6 Earth radius nearby distance at 20 km/s , in an ideal (unique) resonant behavior of the terrestrial oceans. The geological complexity would require a more realistic multi-frequency behavior. The peak energy, in joule unity, is the one for a Moon-like passage at a distance, ten times nearer than our Moon.

## 8. Dedication

This article is dedicated to the memory of victims of past and recent human catastrophes, many of which were driven by hatred, violence, and dehumanizing narratives. A century ago, false accusations and scapegoating contributed to the persecution and extermination of Jews in Europe. Similar waves of hatred can return , as a tsunami, under new forms. Their consequences ultimately threaten not just the Jewish people survival, as in the past, but also the same humanity safety.

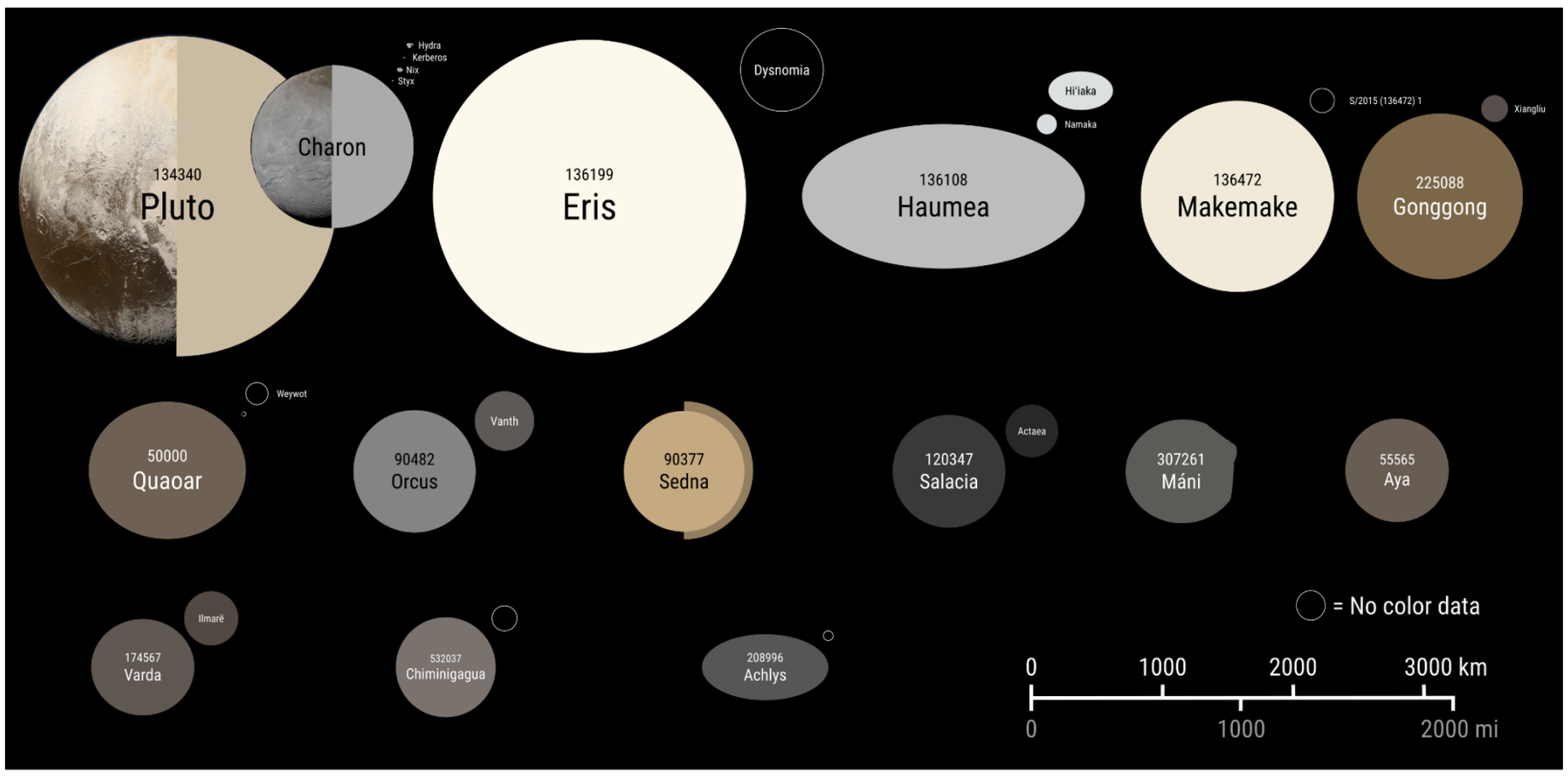

**Figure 16:** Dwarf planet better known, with their name and size. The image of the most recent 2017 OF 201 dwarf planet is not shown here, but it has a size comparable to Sedna diameter.